\documentstyle[aps,epsfig,twocolumn]{revtex}
\begin{document}

\title{\boldmath Forward $K^+$-production in subthreshold
  \boldmath$pA$ collisions at 1.0 GeV}

\author{V.~Koptev,$^1$ M.~B\"uscher,$^2$ H.~Junghans,$^2$
  M.~Nekipelov,$^{1,2}$ K.~Sistemich,$^2$ H.~Str\"oher,$^2$
  V.~Abaev,$^1$ H.-H.~Adam,$^3$ R.~Baldauf,$^4$ S.~Barsov,$^1$
% ...
}

\address{$^1$High Energy Physics Department, Petersburg Nuclear
  Physics Institute, 188350 Gatchina, Russia
}
\address{$^2$Institut f\"ur Kernphysik, Forschungszentrum J\"ulich,
  52425 J\"ulich, Germany
}
\address{$^3$Institut f\"ur Kernphysik, Universit\"at M\"unster,
         W.-Klemm-Str.\ 9, D-48149 M\"unster, Germany
}
\address{$^4$Zentralinstitut f\"ur Elektronik, Forschungszentrum
  J\"ulich, 52425 J\"ulich, Germany
}
\address{$^5$Institut f\"ur Theoretische Physik, Universit\"at
  Gie{\ss}en, H.-Buff-Ring 16, D-35392 Gie{\ss}en, Germany
}
\address{$^6$Institute for Theoretical and Experimental Physics,
  Cheremushkinskaya 25, 117259 Moscow, Russia
}
\address{$^7$High Energy Physics Institute, Tbilisi State University,
  University Street 9, 380086 Tbilisi, Georgia
}
\address{$^8$Institut f\"ur Kern- und Hadronenphysik,
  Forschungszentrum Rossendorf, D-01474 Dresden, Germany
} 
\address{$^9$Laboratory of Nuclear Problems, Joint Institute for
  Nuclear Research, 141980 Dubna, Russia
}
\address{$^{10}$Institut f\"ur Strahlen- und Kernphysik, Universit\"at
         Bonn, Nu{\ss}allee 14, D-53115 Bonn, Germany
}
\address{$^{11}$Institut f\"ur Kernphysik, Universit\"at zu K\"oln,
  Z\"ulpicher Str.\ 77, 50937 K\"oln, Germany
}
\address{$^{12}$Zentralabteilung Technologie, Forschungszentrum
  J\"ulich, 52425 J\"ulich, Germany
}
\address{$^{13}$Institute of Physics, Jagellonian University, Reymonta
  4,30059 Cracow, Poland
}
\address{$^{14}$Dubna Branch, Moscow State University, 141980 Dubna
         Moscow Region, Russia
}
\address{$^{15}$Institute of Nuclear Physics, Radzikowskiego 152,
         PL-31342, Cracow, Poland
}
\address{$^{16}$Physikalisches Institut II, Universit\"at
         Erlangen-N\"urnberg, Erwin-Rom\-mel-Str.\ 1, D-91058 Erlangen, Germany
}
\address{$^{17}$Physics Department, UCL, Gower Street,
         London WC1 6BT, England
}
\address{$^{18}$The Andrzej Soltan Institute for Nuclear Studies, 05400
  Swierk, Poland
} 
\date{\today}
\maketitle

\begin{abstract}
  $K^+$-meson production in $pA$ ($A$ = C, Cu, Au) collisions has been
  studied using the ANKE spectrometer at an internal target position
  of the COSY-J\"ulich accelerator. The complete momentum spectrum of
  kaons emitted at forward angles, $\vartheta\leq 12^{\circ}$, has been
  measured for a beam energy of $T_p=1.0$~GeV, far below the free $NN$
  threshold of 1.58 GeV. The spectrum does not follow a thermal 
  distribution at low kaon momenta and the larger momenta reflect a
  high degree of collectivity in the target nucleus.
\end{abstract}

\narrowtext
A central topic of hadron physics is the influence of the nuclear
medium on elementary processes. This question can be studied by
measuring the production of mesons in nuclei using projectiles with
energies below the threshold for free $NN$ collisions (so-called
subthreshold production). These processes necessarily involve
cooperative effects of the nucleons inside the target nucleus. The
investigation of $K^+$-production is particularly well suited for this
purpose since the meson is relatively heavy so that its production
requires strong medium effects.  Furthermore, the $K^+$ scatters
little in nuclear matter so that its final-state interactions are
expected to be small.

Proton-induced $K^+$-production at subthreshold energies has been
studied at several accelerators. Total cross sections have been
measured at the PNPI synchro-cyclotron for targets between Be and Pb
and projectile energies $T_p$ from 0.8 to 1.0~GeV~\cite{pnpi}. The
results were discussed in terms of different
models~\cite{pnpi,cassing,roc,sibirtsev,paryev}, in particular of
single or two-step reactions involving the creation of an intermediate
pion. It was concluded that additional experimental data were needed
for an unambiguous determination of the reaction mechanism and the
extraction of the information on nuclear-medium effects. Inclusive
differential cross sections for $pA$-interactions have been studied at
BEVALAC~\cite{schnetzer}, SATURNE~\cite{debowski} and
CELSIUS~\cite{badala}. Partial momentum spectra have been obtained at
laboratory emission angles of $10^{\circ}$, $15^{\circ}$,
$35^{\circ}$, $40^{\circ}$, $60^{\circ}$, $80^{\circ}$ and
$90^{\circ}$, at projectile energies down to 1.2~GeV and evidence for
the dominance of the two-step processes has been observed at the
lowest $T_p$. Subthreshold production has also been studied at the
ITEP-synchrotron~\cite{kiselev}, where kaons with high momenta were
identified at projectile proton energies of 1.75 to 2.6~GeV, and in
heavy-ion reactions at GANIL~\cite{ganil} and GSI~\cite{gsi}.

\begin{figure}[ht]
  \vspace*{-5mm}
  \begin{center}
    \leavevmode
    \psfig{file=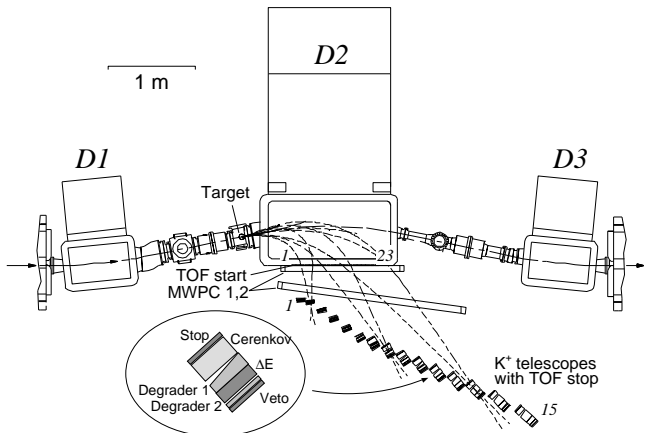,width=8.5cm}
    \caption{Top view of the ANKE spectrometer in the COSY ring and detectors
      used for K$^+$ identification. The inset shows the components of
      one of the range telescopes. Typical kaon trajectories are
      indicated. Details can be found in
      {\protect\cite{ANKE_alt,ANKE_NIM,K_NIM,MESON}}.} 
  \label{fig:anke}
  \end{center}
\end{figure}

The COoler SYnchrotron COSY-J\"ulich~\cite{cosy}, which provides
proton beams in the range $T_p = 0.04 - 2.65$~GeV, is well suited for
the study of K$^+$-production. In measurements with very thin
windowless internal targets, secondary processes of the produced
mesons can be neglected.  Subthreshold $K^+$-production was a prime
motivation for building the ANKE spectrometer~\cite{ANKE_alt,ANKE_NIM}
within one straight section of the COSY ring. It consists of three
dipole magnets D1--D3, see Fig.~\ref{fig:anke}, which separate
forward-emitted reaction products from the circulating proton beam and
allow a determination of their emission angles and momenta. The layout
of the device, including detectors and the data-acquisition (DAQ)
system, was optimized for the study of $K^+$-spectra down to $T_p =
1.0$~GeV. This is a very demanding task because of the small
$K^+$-production cross sections, e.g.\ 39~nb for $p$C collisions at
1.0~GeV~\cite{pnpi}. Kaons have to be identified in a background of
secondary protons and pions which is up to $10^6$ times more intense.
$K^+$-identification is described in detail in~\cite{K_NIM,MESON};
only basic features of the procedures are summarized here.

The COSY ring was filled with $2-4\times 10^{10}$ protons per cycle
and cycle times were typically 30~s. The protons were accelerated to
1.0~GeV on an orbit below the target and then raised by steerers such
that the rate in the ANKE detectors was kept constant and on a level
that could be handled by the DAQ system.  Thin strip targets of C
(polycrystalline diamond), Cu and Au with thicknesses of $40 - 1500\ 
\mu\mathrm g/cm^2$ were used.  Reaction products were detected with an
array of 15 range telescopes positioned along the focal surface. The
time-of-flight (TOF) was measured with 0.5 to 2~mm thick start
scintillators (23 detectors of 5~cm width) and 1~cm thick (10~cm wide)
stop scintillators which are components of the telescopes, see
Fig.~\ref{fig:anke}. Due to the momentum focussing of the dipole D2,
the telescopes can be used to identify different ejectiles via their
ranges: protons are stopped before reaching the $\Delta E$ counters,
$K^+$-mesons come to rest before the veto counters, while pions give
signals in all detectors.  Copper degraders are used to slow down the
kaons in front of the $\Delta E$ counters (degrader 1) and prevent
them from reaching the veto counters (degrader 2). Pions and muons
from the decay of stopped kaons are emitted isotropically, and
partially detected in the veto counter, with a characteristic delay
corresponding to the kaon lifetime $\tau=12.4$~ns. The multiwire
proportional chambers MWPC 1 and 2 allow one to determine the ejectile
tracks and hence separate particles originating from the target from
scattered background, e.g., from the pole shoes of D2.  Applying the
criteria {\em i)} TOF, {\em ii)} energy losses in the scintillation
counters, {\em iii)} delayed signals in the veto counters and {\em
  iv)} track origin in the target, it is possible to identify
unambigously K$^+$-mesons~\cite{K_NIM}. Criteria {\em i)} and {\em
  iii)} were applied on-line to decrease the amount of data written to
tape.

The typical number of identified $K^+$-mesons, $N_{{\mathrm
    tel}(i)}^K$, in one telescope $i$ during a four days run on $C$
was 100.  Since each telescope covers a given momentum range, the
$N_{{\mathrm tel}(i)}^K$ determine directly the momentum distribution.
The accepted momentum bites range from 13\% at the low momentum side
of the telescope array ($\sim 150$~MeV/c) to 7\% for high momenta
($\sim 510$~MeV/c). The vertical angular acceptance of ANKE varies
between $\vartheta_{\mathrm V}=\pm 7^{\circ}$ and $\pm 3.5^{\circ}$
for the low- and high-momentum telescopes, respectively.  In the
horizontal direction, ejectiles with emission angles
$|\vartheta_{\mathrm H}|\leq 12^{\circ}$ were accepted by the on-line
TOF-trigger system~\cite{K_NIM}.

Double differential cross sections for $K^+$-production were obtained
from $N_{{\mathrm tel}(i)}^K$ by comparison with the number of pions
$N_{{\mathrm stop}(i)}^\pi$ detected in the stop counters of the same
telescope, which were measured in dedicated calibration runs.  The
ratio of $K^+$ and $\pi^+$ counts was corrected for the different
detection efficiencies $\epsilon$ and normalized to the corresponding
proton-beam fluxes $M^{\pi}$ and $M^{K}$ (i.e.\ luminosities) and to
the cross sections of $\pi^+$ production in $p$C interactions in the
forward direction at 1.0~GeV~\cite{Coch,Papp,Abaev}. Due to their much
larger production cross section, pions can be identified applying only
criteria {\em i)} and {\em iv)}.  The cross section is thus given by:
\begin{eqnarray}
  \label{eq:cross-section}
  \frac{d ^2 \sigma^K}{d\Omega dp} = \frac{d^2\sigma^\pi}{d\Omega dp} \cdot
  \frac{N^K_{{\mathrm tel}(i)}}{N^\pi_{{\mathrm stop}(i)}} \cdot
  \frac{M^\pi}{M^K} \cdot \frac{1}{\epsilon} \\
  \epsilon = 
    \frac{
    \epsilon_{{\mathrm tel}(i)}^K \cdot \epsilon_{{\mathrm MWPC}(i)}^K \cdot
    \epsilon_{{\mathrm decay}}^{K}} 
    {\epsilon_{{\mathrm stop(i)}}^\pi \cdot \epsilon_{{\mathrm
    MWPC}(i)}^\pi \cdot \epsilon_{\mathrm decay}^\pi} \nonumber \ ,
\end{eqnarray}
where $\epsilon^K_{{\mathrm tel}(i)}$ is the $K^+$-identification
efficiency of the $i\,^{\mathrm th}$ telescope including the detection
probability for the decay muons and pions. Only events where veto
counter signals were delayed by at least 2.5~ns with respect to those
of the stopped kaons were considered. The efficiencies,
$\epsilon^K_{{\mathrm tel}(i)}\sim 0.1-0.3$, were obtained from a
calibration run at $T_p=2.3$~GeV, where $K^+$-production is
significantly larger~\cite{K_NIM}.

The average pion detection efficiencies $\epsilon_{{\mathrm
    stop}(i)}^\pi= 0.98$. The MWPC efficiencies $\epsilon_{\mathrm
  MWPC}$ are in the range 0.97--0.99 both for pions and kaons.
$\epsilon_{\mathrm decay}^K$ and $\epsilon_{\mathrm decay}^\pi$ take
into account losses in flight between target and stop counters; they
were determined by simulations as 0.30--0.36 for kaons and 0.85--0.88
for pions.  Relative monitoring of the proton beam interacting with
the target nuclei $A$ ($M_A^{K}$ and $M_A^{\pi}$) was done with an
accuracy of 2\% using 4-fold coincidences of telescopes 2, 3, 4 and 5
(selection of ejectiles by-passing the spectrometer dipole D2).

\begin{figure}[ht]
\vspace*{-6mm}
  \begin{center}
    \psfig{file=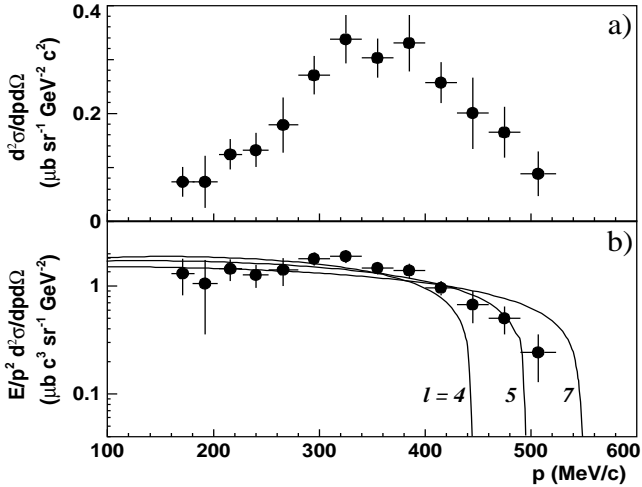,width=8.5cm,height=6.5cm}
    \caption{a) Double differential $K^+$-production cross section 
      for the 
      $p(1.0\ {\mathrm GeV})^{12}C \rightarrow K^+(\vartheta \leq
      12^{\circ})X$ reaction as a function of the $K^+$-momentum. 
      b) Same data plotted as invariant cross section. The error bars 
      are purely statistical. The overall normalization uncertainty
      is estimated to be 10\%. The solid
      lines describe the behavior of the invariant cross section
      within a phase-space approximation (Eq.~(\ref{eq:kinem})).}
  \label{fig:1.0GeV}
  \end{center}
\end{figure}

Our double differential cross section for $K^+$ production in
proton-carbon interactions at $T=1.0$~GeV is shown in
Fig.~\ref{fig:1.0GeV}a). In contrast to measurements at higher
energies~\cite{schnetzer,debowski,badala,kiselev,buescher}, ANKE
reveals, for the first time, a complete momentum spectrum at deep
subthreshold energies.

The {\em low-momentum} part of the $K^+$-spectrum in
Fig.~\ref{fig:1.0GeV}b) cannot be described by a thermal distribution,
$E/p^2\, d^2\sigma/dp\,d\Omega\propto\exp(-T^{\star}/T_0)$, where
$T^{\star}$ is the $K^+$ kinetic energy in the beam-proton nucleus
system. Such distributions had been assumed to deduce total cross
sections from earlier measurements with limited momentum
intervals~\cite{schnetzer,debowski,badala,buescher}.

A high degree of collectivity is needed in the target nucleus to allow
$K^+$-production far below the free NN-threshold, i.e.\ the number of
target nucleons involved must be significantly larger than
one~\cite{pnpi,cassing,roc,sibirtsev,paryev}. Alternatively, high
intrinsic momenta of the participating target nucleon(s) are required
to supply the missing energy for subthreshold kaon production. This is
particularly the case for the {\em high-momentum} part of the kaon
spectrum: internal momenta of a single nucleon of at least $p_N \sim
350(550)$~MeV/c would be needed in order to produce kaons in the
forward direction with momenta of $p_K\sim 260(500)$~MeV/c.  High
momentum components above $\sim 500$~MeV/c are essentially due to
many-body correlations in the nucleus \cite{correlations}.

To get a rough estimate of the number of participating nucleons, we
describe the invariant cross section within a phase-space
approximation. This method has previously been applied to
$K^+$-production data~\cite{buescher} in order to compare spectra
obtained under different kinematical
conditions~\cite{schnetzer,debowski,buescher}. The invariant cross
section for the $p{+}(lN){\to}(lN){+}\Lambda{+}K^+$ reaction is then
\begin{equation}
  E\frac{d^3\sigma}{dp^3} \propto
  \frac{\sqrt{(s_l-m_\Lambda^2 - l^2m_N^2)^2-4m_\Lambda l^2m_N^2}}{s_l}
  \label{eq:kinem}
\end{equation} 
where $m_\Lambda$ and $m_N$ are the $\Lambda$ and nucleon masses,
respectively. $l$ is the number of nucleons involved in the
interaction and
\begin{equation}
  s_l{=}s{+}m_K^2{-}2E_K(T_p{+}[l{+}1]m_N){+}2p_Kp_pcos\theta_K\ .
\end{equation}
$m_K$, $E_K$, $p_K$ and $\theta_K$ are the kaon mass, total energy,
momentum and emission angle, respectively, while $T_p$ and $p_p$
denote the beam energy and momentum.  $s$ is the square of the CM
energy of the incident proton and the $l$ target nucleons. The solid
lines in Fig.~\ref{fig:1.0GeV}b) show the momentum dependence of the
invariant cross section from Eq.~(\ref{eq:kinem}) for $l=4,5,7$.
Although this neglects the intrinsic motion of the $l$ target
nucleons, it shows that kaon production at $T=1.0$~GeV can only be
understood in terms of cooperative effects with the effective number
of nucleons involved in the interaction being $\sim5-6$. It has been
suggested \cite{pnpi,cassing,roc,sibirtsev,paryev} that such
cooperative effects can be described in terms of multi-step mechanisms
or high-momentum components in the nuclear wave function.

It has also been pointed out~\cite{pnpi,cassing,roc,sibirtsev,paryev}
that the mechanism of subthreshold $K^+$ production might be
identified from the $A$-dependence of the production cross section.
The total cross sections measured at PNPI in the energy range
$T=0.8-1.0$~GeV scale as $\sigma_{\mathrm tot} \propto
A^1$~\cite{pnpi}. This behavior has been interpreted in terms of a
two-step mechanism with the formation of an intermediate $\pi$-meson.

\begin{figure}[htbp]
\vspace*{-6mm}
  \begin{center}
    \leavevmode
    \psfig{file=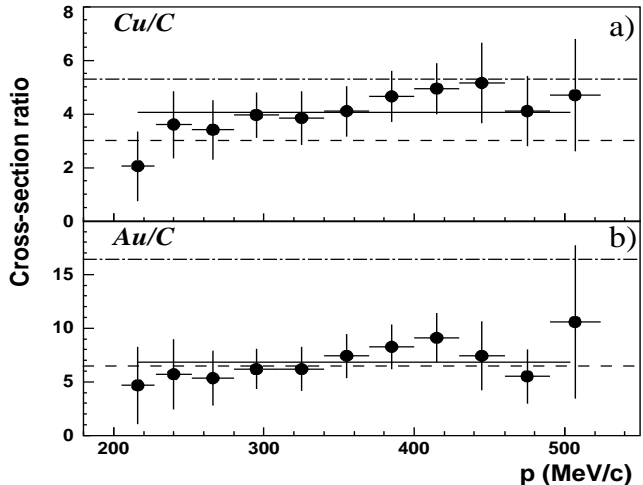,width=8.5cm,height=6.5cm}
    \caption{Ratios of the $K^+$ production cross sections for Cu/C
      and Au/C as a function of the kaon momentum. The solid
      horizontal lines indicate a fit by a constant value. The dashed
      lines illustrate the ratios if the cross sections scale
      like $A^{2/3}$ (dashed) and $A^1$ (dashed-dotted).}
    \label{fig:A-1.0GeV}
  \end{center}
\end{figure}

We used targets of widely different masses to determine the
$A$-dependence of the differential cross sections. All measurements
were carried out with the same geometry and detection system at ANKE
and the same proton beam settings. The ratios of the kaon-production
cross sections for various nuclei as a function of kaon momentum
$\sigma_A^{K^+}/\sigma_C^{K^+}$ are therefore equal to the
corresponding ratios of kaon count rates in the individual telescopes,
normalized to the pion cross-section ratio at a momentum of $507\pm
17$~MeV/c detected in telescope \#15:
\begin{equation}
  \left(\frac {\sigma_A^{K}}{\sigma_C^{K}}\right)_{\!\!(i)}\!\!=\!
  \left(\frac{N_A^{K}}{N_C^{K}}\right)_{\!\!(i)}
  \frac{M_C^{K}}{M_A^{K}}
  \left(\frac{N_C^{\pi}}{N_A^{\pi}}\right)_{\!\!(15)}
  \frac{M_A^{\pi}}{M_C^{\pi}}
  \left(\frac{\sigma_A^{\pi}}{\sigma_C^{\pi}}\right)_{\!\!(15)}\!\!.
\end{equation}
The cross-section ratios for producing pions with momenta around
500~MeV/c in the forward direction in $pA$ collisions were measured to
better than 10\% by several groups in the $0.73 - 4.2$~GeV energy
range~\cite{Coch,Papp,Abaev,MESON}. All uncertainties from the
efficiency correction $\epsilon$ in Eq.~(\ref{eq:cross-section})
cancel out for such ratios. The results for Cu/C and Au/C are shown in
Fig.~\ref{fig:A-1.0GeV}.

The ratios are almost independent of the $K^+$-momentum. The solid
lines in Fig.~\ref{fig:A-1.0GeV} indicate fits with constant values;
$R($Cu/C$)=4.0\pm 0.3$ and $R($Au/C$)=6.8\pm0.8$. These values should
be compared to the expected ratios if the cross sections scaled as
$A^1$: $R($Cu/C$)\sim 5$ and $R($Au/C$)\sim 16$.  Our data show a
significantly weaker $A$-dependence with an exponent closer to 2/3.
This is in contrast to the behavior of the total cross sections at the
same beam energy found in~\cite{pnpi}, where kaon production was
described in terms of the two-step mechanism.  It is possible that,
due to rescattering of the produced kaons in the target nucleus out of
the angular acceptance of ANKE, there is a kaon deficiency at high
momenta and small angles.  Since this effect should be stronger for
heavier targets, it might account for the weaker $A$-dependence of the
differential cross section.

Summarizing, we have measured small-angle $K^+$-production at
$T=1.0$~GeV, which is far below the free NN threshold. The
low-momentum part of the $K^+$-spectrum does not follow a thermal
distribution whereas the high-momentum part reveals a high
collectivity of the target nucleus. Within a simple kinematical model,
about 5--6 nucleons are needed to allow kaon production with momenta
$p_K\sim500$~MeV/c.  Alternatively, intrinsic momenta of at least
$p_N\sim 550$~MeV/c are required if such kaons are produced in a
collision with a single target nucleon. The target-mass dependence of
the differential cross sections shows a scaling significantly weaker
than those of the total cross sections~\cite{pnpi}. The latter have
been interpreted in terms of two-step $K^+$-production. It remains to
be shown by microscopic calculations whether this finding is due to
rescattering effects of the kaons in the target nucleus or whether it
is a reflection of single-step kaon production with high-momentum
components in the nuclear wave function.

We would like to acknowledge the assistance we received in performing
these measurements at the ANKE spectrometer.  Financial support from
the following funding agencies was of indispensable help for building
ANKE, its detectors and DAQ: Georgia (Department of Science and
Technology), Germany (BMBF: grants WTZ-RUS-649-96, WTZ-RUS-666-97,
WTZ-RUS-685-99, WTZ-POL-007-99; DFG: 436 RUS 113/337, 436 RUS 113/444,
436 RUS 113/561, State of North-Rhine Westfalia), Poland (Polish State
Committee for Scientific Research: 2 P03B 101 19), Russia (Russian
Ministry of Science, Russian Academy of Science: 99-02-04034,
99-02-18179a) and European Community (INTAS-98-500).


\begin{references}
\bibitem[*]{nowat}  Now working at: Saint-Gobain Crystals \& Detectors
  GmbH, 42929 Wer\-mels\-kir\-chen, Germany.
%
\bibitem{pnpi}V.P.~Koptev {\em et al.}, JETP {\bf 67}, 2177
  (1988).
%
\bibitem{cassing}W.~Cassing {\em et al.}, Phys.\ Lett.\ B {\bf 238}, 25
  (1990).
%
\bibitem{roc}H.~M\"uller and K.~Sistemich, Z.\ Phys.\ A {\bf 344},
  197 (1992)
%
\bibitem{sibirtsev}A.A.~Sibirtsev and M.~B\"uscher, Z.\ Phys.\ A {\bf
    347}, 191 (1994) 
%
\bibitem{paryev}E.Ya.~Paryev, Eur.\ Phys.\ J.\ A {\bf 5}, 307
  (1999).
%
\bibitem{schnetzer}S.~Schnetzer {\em et al.}, Phys.\ Rev.\ C {\bf 40}, 640
  (1989). 
%
\bibitem{debowski}M.~Debowski {\em et al.}, Z.\ Phys.\ A {\bf 356},
  313 (1996).
%
\bibitem{badala}A.~Badal\`{a} {\em et al.}, Phys.\ Rev.\  Lett.\ {\bf 80},
  4863 (1998). 
%
\bibitem{kiselev}Yu.T.~Kiselev {\em et al.}, J.\ Phys.\ B {\bf 25}, 381 (1999).
%
\bibitem{ganil}J.~Julien {\em et al.}, Phys.\ Lett.\ B {\bf 264}, 269 (1991).
%
\bibitem{gsi}P.~Senger and H.~Str\"obele, J.\ Phys.\ G {\bf 25}, R59
  (1999). 
%
\bibitem{cosy}R.~Maier,  Nucl.\ Instr.\ Methods Phys.\ Res., Sect.\ A 
   {\bf 390}, 1 (1997). 
%
\bibitem{ANKE_alt}M.~B\"uscher {\em et al.}, Phys.\ Scripta RS 
  {\bf 21}, 23 (1993).
%
\bibitem{ANKE_NIM}S.~Barsov {\em et al.}, Nucl.\ Instr.\ Methods Phys.\ Res.,
  Sect.\ A {\bf 462/3}, 364 (2001).
%
\bibitem{K_NIM}M.~B\"uscher {\em et al.}, Nucl.\ Instr.\ Methods Phys.\ Res.,
  Sect.\ A (submitted), available via:
  http://ikpd15.ikp.kfa-juelich.de:8085/doc/Publications.html.
%
\bibitem{MESON}S.~Barsov {\em et al.}, Acta Phys.\ Polonica B {\bf 31}, 2159
  (2000). 
%
\bibitem{Coch}D.R.F.~Cochran {\em et al.}, Phys.\ Rev.\ D {\bf 6}, 3085
  (1972). 
%
\bibitem{Papp}J.~Papp {\em et al.}, Phys.\ Rev.\ Lett.\ {\bf 34}, 601 (1975).
%
\bibitem{Abaev}V.V.~Abaev {\em et al.}, J.\ Phys.\ G {\bf 14}, 903 (1988).
  {\bf 31}, 2159 (2000).
%
\bibitem{buescher}M. B\"uscher et al., Z.\ Phys.\ A {\bf 355}, 93 (1996). 
%
\bibitem{correlations}H.~Nifenecker, J.A.~Pinston, Prog.\ Part.\
  Nucl.\ Phys.\ {\bf 23}, 271 (1989).
\end{references}
\end{document}